\documentclass[reprint,amsmath,aip]{revtex4-2}

\usepackage{graphicx}
\usepackage{dcolumn}
\usepackage{bm}
\usepackage{booktabs, multirow} 
\usepackage{soul}
\usepackage[table]{xcolor} 
\usepackage{changepage,threeparttable} 
\usepackage{multirow}
\usepackage{amsmath}
\usepackage{amssymb}
\usepackage{amsfonts}

\begin{document}

\preprint{APS/123-QED}

\title{Polymer-loaded three dimensional microwave cavities for hybrid quantum systems}

\author{M.~Ruether}

\author{C.A.~Potts}%
\author{J.P.~Davis}
\email{jdavis@ualberta.ca}
\author{L.J.~LeBlanc}
\email{lindsay.leblanc@ualberta.ca}
\affiliation{%
 Department of Physics, University of Alberta, Edmonton, Alberta T6G 2E1, Canada}%

\date{\today}

\begin{abstract}
Microwave cavity resonators are crucial components of many quantum technologies and are a promising platform for hybrid quantum systems, as their open architecture enables the integration of multiple subsystems inside the cavity volume.  To support these subsystems within the cavity, auxiliary  structures are often required, but the effects of these structures on the microwave cavity mode are difficult to predict due to a lack of \textit{a priori} knowledge of the materials' response in the microwave regime.  Understanding these effects becomes even more important when frequency matching is critical and tuning is limited, for example, when matching microwave modes to atomic resonances. Here, we study the microwave cavity mode in the presence of three commonly-used machinable polymers, paying particular attention to the change in resonance and the dissipation of energy.  We demonstrate how to use the derived dielectric coefficient and loss tangent parameters for cavity design in a test case, wherein we match a polymer-filled 3D microwave cavity to a hyperfine transition in rubidium.
\end{abstract}

\maketitle

\section{Introduction}
The high-quality factors and open architecture of three dimensional (3D) microwave cavities \cite{Turneaure68,Reagor13,Suleymanzade20,Chakram20,Kudra20,Sharping20,McRae20} has enabled hybridization with a wide range of classical and quantum systems, including 3D transmon qubits \cite{Vlastakis13,Flurin15,Lane20}, magnonic resonators \cite{Huebl13,Tabuchi14,Zhang14,Goryachev14,Bai15,Potts20a,Potts20b}, neutral atoms and molecules \cite{Goy83,Stammeier17,Wright19,Adwaith19,Suleymanzade20,Tretiakov20}, quantum dots \cite{Kong15}, and piezoelectric optomechanical resonators \cite{Ramp20}.  In most of these systems, it is important that --- due to a resonant interaction --- the microwave cavity frequency be identical to that of the desired transition in the system of interest: for example, the electronic transitions in atoms \cite{Tretiakov20}, or the mechanical modes in piezo-optomechanics \cite{Ramp20}.  While in some scenarios, the hybrid system is easily tunable over a wide range, such as the magnon frequencies in cavity magnon polaritons \cite{Tabuchi14}, these are the exception and generally, the tunability is limited \cite{Tretiakov20} or zero \cite{Reagor13}.  Hence, to achieve resonant interaction and efficient coupling, all components of the 3D microwave cavity must be precisely understood at the design stage to accurately produce a microwave cavity with the desired frequency.  This includes any materials inserted into the cavity, often machinable polymers, used to physically support auxiliary components.  

To date, our attempts to use standard values for the dielectric constant of common machinable polymers have proven to have insufficient accuracy to properly design resonant microwave cavities.  Here, we report frequency-dependent values of the dielectric constant for three common machinable polymers, as well as the relative microwave cavity loss due to these materials.  These measurements will assist those seeking to construct resonant microwave cavities, especially in hybrid systems with limited tunability.  

\begin{figure}[b]
\includegraphics[scale=0.9]{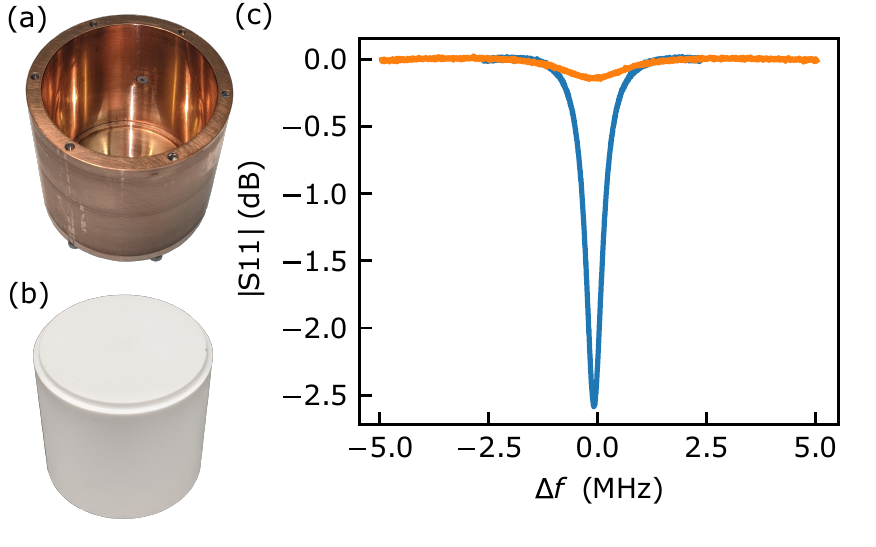}
\caption{\label{fig:epsart} a) The microwave test cavity with the top end-cap removed. The cavity internal dimensions are a diameter of 58.07~mm, and height 57.04~mm with a 2$\times$2~mm step on the end-caps. b) A polymer plug machined to match the inside of the cavity to a snug fit. The step is used to separate otherwise degenerate modes from the TE$_{011}$ resonance \cite{Reagor13}. c) $S_{11}$ measurements of the TE$_{011}$ mode, in air (blue) and PTFE (orange), plotted as the detuning from cavity resonance, showing the change in the external coupling with the addition of the PTFE.}
\end{figure}

\section{Methods}
In this work, we focus on measurements of three polymer materials, polytetrafluoroethylene (PTFE), polypropylene (PP), and polyethylene (PE).  [Note that we also tested an additional six materials (nylon, anti-static polyethylene, moisture-resistance polyethylene, polyester, polycarbonate, and Garolite LE), but we found that these suffered from considerable dielectric loss at GHz-frequencies and therefore made microwave modes generally unobservable.]  We determine the effective energy loss of several cavity modes for the three materials by comparing cavity measurements to finite-element simulations. Using a cylindrical copper cavity similar to those in Refs.~\cite{Reagor13,Tretiakov20}, we measure the frequencies and internal quality factors of multiple modes, both for air-filled cavities,  and for cavities with tight-fitting polymer plugs, as shown in Fig.~1. Microwave reflection ($S_{11}$) measurements are performed on a vector network analyser (VNA). In order to couple microwaves into and out of the cavity, an SMA pin-coupler acted as a waveguide mounted at the midpoint of the cylindrical axis and offset 15~mm tangentially; this was modified to be flush with the cavity wall so the entire void of the cavity is filled with material. 

\section{Dielectric Constant}
We describe the dielectric properties of the polymers using the Drude-Lorentz oscillator model~\cite{Fox}. In this model, the electrons within the material are treated as damped harmonic oscillators. Therefore, due to the polarizability of the material, the dielectric constant can be written as a function of the material's complex susceptibility, with $\varepsilon_{\rm{r}} = 1 + \tilde{\chi}$ and \cite{Pozar}
\begin{equation}
    \varepsilon_{\rm{r}} = \varepsilon^{\prime}+i \varepsilon^{\prime\prime},
\end{equation}
where the real part $\varepsilon^{\prime}$ is associated with the index of refraction of the material, and the imaginary part $\varepsilon^{\prime\prime}$ describes the energy loss due to absorption.

To study the microwave-frequency dielectric response of various materials, we measure the properties of resonant modes of a microwave cavity with and without a dielectric filling the interior volume.  The resonance frequencies of the transverse electric (TE) modes of a cylindrical cavity filled with a dielectric are
\begin{align}
    f_{nm\ell} = \frac{c}{2 \pi \sqrt{\mu_r \varepsilon^{\prime}}} \sqrt{\left(\frac{p^{\prime}_{nm}}{a}\right)^2+\left(\frac{\ell \pi}{d}\right)^2},
\end{align}
where $f_{nm\ell}$ is the frequency of the mode with indices $(n,m,\ell)$; $c$ is the speed of light in  vacuum; $a$ is the internal radius of the cavity; $d$ is the internal height; and $\mu_r$ is relative permeability (for non-magnetic materials,  $\mu_r=1$). The cavity mode is defined through the Bessel function parameter $p'_{nm}$, where $p'_{nm}$ is the \textit{m}$^{\rm th}$ root of $J'_{n}$ \cite{Pozar}.  We emphasize that to study the effect of changing materials, and thus $\varepsilon'$, we must consider the same mode (with indices $(n,m,\ell)$) under both air- and polymer-filled conditions. 

When transitioning between an empty cavity and one filled with dielectric, tracking the resonances can be difficult due to mode-dependent changes in the external coupling, in addition to changes in their resonance frequencies. In order to ensure proper mode identification, a combination of preliminary finite element simulations using approximate values of the dielectric constant, as well as tracking mode frequencies relative to the highest $Q$ mode (the $\textrm{TE}_{011}$ mode), was performed. 

Using this tracking, we measure and identify the mode frequencies and quality factors for 7-10 cavity resonances in each material (Table~\ref{tab:allData}), then extract  values of $\varepsilon'$ calculated from Eq.~2, as shown in Fig.~\ref{fig:epsart}. A linear fit well represents the frequency dependence of the dielectric constant in this range, and allows for the extrapolation to frequencies needed for novel designs.  From the fit results, the frequency dependence of the dielectric constants are tabulated in Table~\ref{tab:slope}.  In the case of PE, the frequency-dependence is null, within the uncertainty of the fit, which may have advantages in some experimental designs.  The choice of dielectric material will influence cavity designs, with higher-dielectric materials allowing for smaller cavities and greater mode confinement, while lower-dielectric materials allow for the design of larger cavities to accommodate particular experiments.  

\begin{figure}
\includegraphics[scale=0.49]{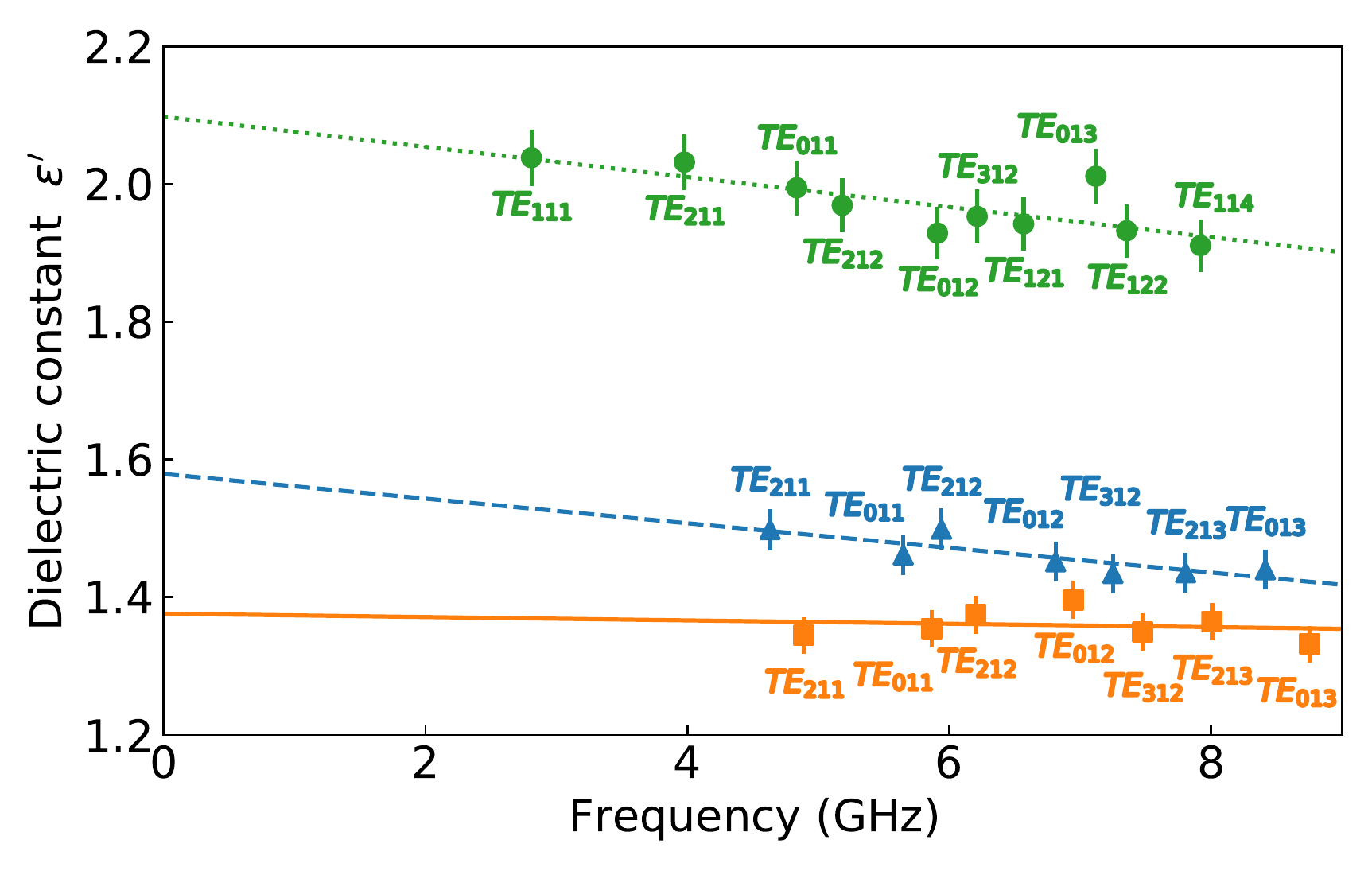}
\caption{\label{fig:epsart} Real part of the relative dielectric constant, $\varepsilon'$, for each of the labeled modes. Green circles are for polytetrafluoroethylene (PTFE), blue triangles are for polypropylene (PP), and orange squares are for polyethylene (PE). The uncertainties shown represent a 0.02\% variance \cite{Jacob02} in the material's dimensions due to machining tolerances.}
\end{figure}

\begin{table}[!htp]\centering
\footnotesize	
\begin{tabular}{l r r}
\hline\hline
Material & \multicolumn{1}{c}{$d\varepsilon^\prime/df$} & \multicolumn{1}{c}{$\varepsilon_{\rm r,0}$}\\
&  \multicolumn{1}{c}{(GHz$^{-1}$)} & \\
\hline
PTFE & -0.022(6) & 2.10(4) \\
PP & -0.017(5) & 1.58(3)\\
PE & -0.002(7) & 1.37(5)\\ 
\hline\hline
\end{tabular}
\caption{Frequency-dependence of the dielectric constant $\varepsilon^\prime$, where $\varepsilon^\prime(f)  = (d\varepsilon^\prime/df) f + \varepsilon_{\rm r,0}$.  Parameters are extracted from fits shown  in Fig.~2; parameter uncertainties are statistical and represent one standard deviation.}\label{tab:slope}
\end{table}

{\renewcommand{\arraystretch}{1.2}}
\begin{table}[!htp]\centering
\begin{tabular}{llrrrrr}

\hline
\hline
Mode \quad & Material & \multicolumn{1}{c}{$Q_i$}  & \multicolumn{1}{c}{$f$} & \quad $G_{\rm seam}$ \quad  & \multicolumn{1}{c}{$\varepsilon'$} &  \multicolumn{1}{c}{$\tan \delta$}\\
 &  &  &  \multicolumn{1}{c}{(GHz)} &   &  &  $(10^{-3})$\\
\hline

 \multirow{4}{2em}{$\textrm{TE}_{211}$} &
 Air & 9805 &\quad5.67542 &\quad1.44306 &  &  
 \\& PTFE & 926 &3.97692 &  &2.0322 &\quad0.985 
 \\& PP &2158 &4.63122 &  &1.4985 &0.2804
 \\& PE &1044 &4.8892 & &1.3446 &0.2716 \\
\hline

 \multirow{4}{2em}{$\textrm{TE}_{011}$} & Air &27635 &6.82211 &5.12826 &  &  
 \\ & PTFE &2382 &4.83447 &  &1.9949 &0.3857 \\ & PP &4855 &5.64805  &  &1.4616 &0.1334 
 \\ & PE &2973 &5.86669  &  &1.3547 &0.1234  
 \\\hline

\multirow{4}{2em}{$\textrm{TE}_{212}$} 
&Air &4485 &7.29424 &2.87271 &  & 
\\ & PTFE&1955 &5.18215 & &1.9694 &0.4414 
\\ & PP & 3972 &5.93977 &  &1.499 &0.0978 
\\ & PE &1934 &6.20285 &  &1.3746 &0.0706   \\\hline

\multirow{4}{2em}{$\textrm{TE}_{012}$}
& Air &33621 &8.20274 &4.46773 &   &   
\\ & PTFE &3529 &5.90971 &  &1.929 &0.2577 
\\&  PP &3396 &6.81213&   &1.4518 &0.1969 
\\& PE & 3153 &6.94666 &  &1.3961 &0.1917 
\\\hline

\multirow{4}{2em}{$\textrm{TE}_{312}$} & 
Air &7283 &8.70811 &2.55954  &   &   
\\ & PTFE&3331 &6.21268 & & 1.9533  &0.228 
\\&  PP& 1614 &7.24901 & & 1.4347   &0.2513 
\\& PE&2948 &7.47494 & & 1.3493   &0.3075 
\\\hline

\multirow{4}{2em}{$\textrm{TE}_{013}$} & 
Air &22458 &10.0897 &3.39231 &   &   
\\ & PTFE &3400 &7.11881 &  &2.0117 &0.2129 
\\ & PP & 3791 &8.41341 &  &1.4402 &0.1709 
\\ & PE &1488 &8.74971 &  &1.3317 &0.1231 
\\ \hline \hline
\end{tabular}
\caption{Experimental quality factors and resonance frequencies of the 3D cavity measured in air, PTFE, PP, and PE.  For the air-only cavities, the seam conductivity ($G_{\textrm{seam}}$) is calculated, and for loaded cavities, the real part of the dielectric constant $\varepsilon'$ and dielectric loss tangent (tan $\delta$) are calculated.}\label{tab:allData}
\end{table}

\section{Dissipation}

\begin{figure}[t]
\includegraphics[scale=0.9]{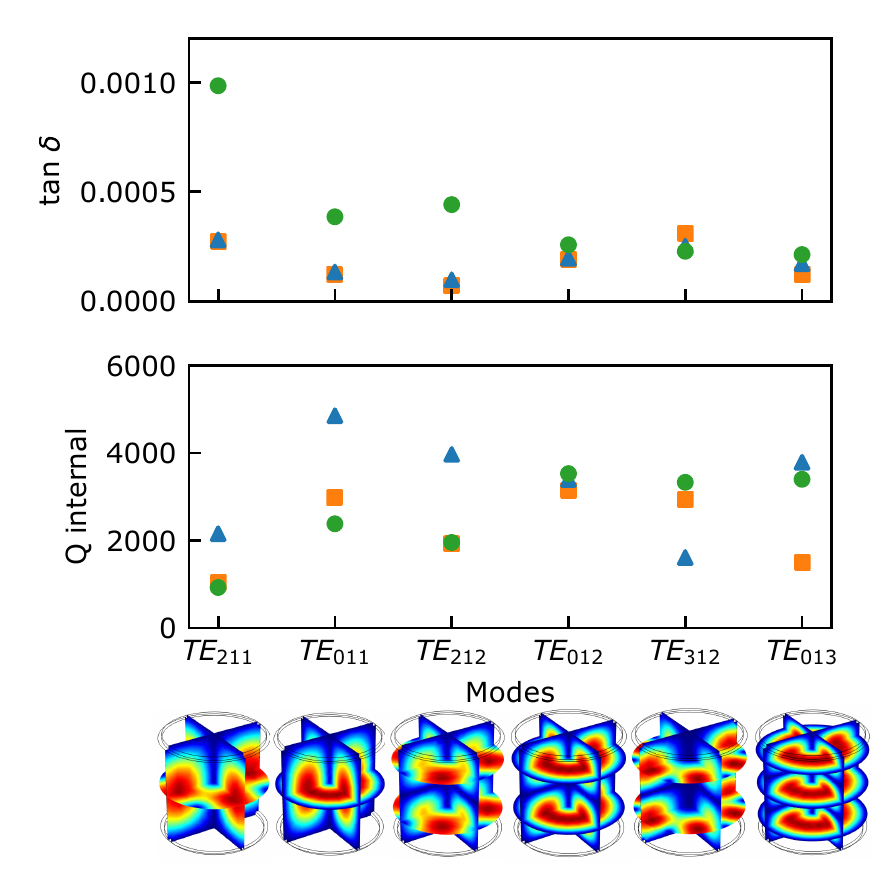}
\caption{\label{fig:epsart} Loaded-cavity dielectric loss tangents for PTFE (green circles), PP (blue triangles) and PE (orange squares) for each of the measured resonances, with a corresponding simulation of the mode shape.  The corresponding mode frequencies can be found in Table 2.}
\end{figure}

The quality factor of a resonator depends on the stored energy compared to the rate at which energy is dissipated. An $S_{11}$ measurement can be fit to extract the internal quality factor of the cavity ($Q_{\textrm{i}}$) and the external quality factor ($Q_{\textrm{ext}}$) \cite{Probst15}:
\begin{equation}
    Q_{\textrm{total}} = \left(\frac{1}{Q_{\textrm{i}}} + \frac{1}{\mathbf{Re}(Q_{\textrm{ext}})} \right)^{-1}.
\end{equation}
$Q_{\textrm{ext}}$ is a measure of the external coupling into and out of the cavity and in general is complex valued due to impedance mismatch, whereas $Q_i$ is the relevant value to measure the cavity performance. The parameter $Q_i$ accounts for loss due to the resistance of the conducting walls of the cavity (associated with $Q_{\textrm{conductor}}$), as well as the dissipation of energy in the dielectric inside the cavity ($Q_{\textrm{dielectric}}$), and the additional resistance across the seam between the cavity end-caps and walls ($Q_{\textrm{seam}}$), such that
\begin{equation}
    Q_{\textrm{i}} = \bigg(\frac{1}{Q_{\textrm{conductor}}} +  \frac{1}{Q_{\textrm{dielectric}}} +
    \frac{1}{Q_{\textrm{seam}}}\bigg)^{-1}.
\end{equation}

Difficulty arises when comparing losses in materials across resonance modes as the effects of the varying mode volumes and external coupling can cause adjustments to the other loss channels. For example, a higher-dielectric material may cause a resonance to be localized in the center of a cavity and therefore have less interaction with the walls and have lower conductor-induced losses. COMSOL has the benefit of accounting for these differences on a case by case basis: if the material conditions are determined by replicating the empty cavity in simulation, then each polymer's dielectric losses can be solved to match the measured $Q_{\textrm{i}}$.
 
In COMSOL, a material's dielectric properties can be introduced using equation (1), where $\varepsilon'$ shifts the resonance frequency  and  $\varepsilon''$ determines dielectric losses. Solving for the cavity-only value of $\varepsilon''$ across materials and resonance modes is  valid only if the experimental quality factor for the cavity without the material can be fully simulated. The two dominant loss channels for the empty cavity are the conductivity of the bulk copper and seam losses between the end caps and the cylindrical body \cite{Pozar,Brecht15}. COMSOL accounts for the resistance in the bulk material in its native quality factor ($\texttt{emw.Qfactor}$) calculations determined by the selected materials bulk conductivity $\sigma$. Losses at the seam must be manually added to the calculations. The quality factor for the seam can be calculated using the line integral form \cite{Brecht15},
\begin{equation}
    \frac{1}{Q_{\textrm{seam}}} = \frac{1}{G_\textrm{seam}} \left( \frac{L \oint_\textrm{seam} | \vec{J}_s \times \hat{l}|^2 dl}    {\omega \mu_0 \oint_\textrm{tot}|\vec{H}|^2 dV}\right).
\end{equation}
Here, $G_{\textrm{seam}}$ is the conductivity across the seam, the numerator $L \int_\textrm{seam} | \vec{J}_s \times \hat{l}|^2 dl$ is the energy loss due to the seam in the conductor, with $L$ as the length of the seam along which the line integral is taken. $J_s$ is the surface current across the seam, $\omega = 2\pi f$ is the angular frequency of the resonance, and $\mu_0 \int_\textrm{tot}|\vec{H}|^2 dV$ is the volume integral of the magnetic field $\vec{H}$ inside the cavity --- to yield the stored energy, with $\mu_0$ as the vacuum permeability.

In order to implement equation (5) into COMSOL, as it lacks cross product functionality, the value for $\vec{J}_s \times \hat{l}$ is replaced with $\texttt{emw.Jy}$, as the $y$-component is the direction of the surface current --- the only one crossing the seam in this configuration. Here, the $y$-component is defined as the axis concentric with the cylinder. In order to faithfully compare COMSOL solutions across simulations, additional modifications included normalizing the surface currents and magnetic fields by the global maximum found in the solution. Using a value for the bulk copper conductivity of $5.89 \times 10^7 \ \Omega \textrm{m}^{-1} $ \cite{Clegg03}, a parametric sweep of the conductivity of $G_{\textrm{seam}}$ can be used to solve for the quality factor of the empty cavity for each of the modes, and hence determination of $G_{\textrm{seam}}$ by comparison with the measured quality factor. 

Equipped with the bare-cavity loss parameters, the additional dissipation due to the dielectric materials can be added to the simulation. For each mode, the unique value for $\varepsilon'$ and $G_{\textrm{seam}}$ are used, and a parametric sweep of $\varepsilon''$ in COMSOL allows determination of the actual $\varepsilon''$ by comparison with the measured dielectric-loaded quality factor. Converting $\varepsilon''$ to an effective $\tan\delta$ can then be calculated via \cite{Pozar,McRae20}
\begin{equation}
    \tan\delta = \frac{\varepsilon''}{\varepsilon'}=\frac{1}{Q_{\textrm{dielectric}}}.
\end{equation}
The resulting effective $\tan\delta$ (loss) values are compared across modes and materials in Fig.~3 and Table~2. Note, for example, that while PTFE is commonly used in microwave cavities, PP and PE are generally lower loss and therefore may be a better candidates for some applications. 

\section{Test Case: Cavity Design} 
As a test of the accuracy and usefulness of these findings, we set out to design a `science' cavity.  The target was to hold a sealed $^{85}$Rb vapor cell for microwave coupling to the 3.035732~GHz hyperfine transition.  Additionally, the cavity was to be made from aluminum and filled with PTFE to cradle the Rb vapor cell.  Note that to match this lower frequency to the high-$Q$ TE$_{011}$ mode, this cavity is considerably larger than an empty cavity, at a final 96.5-mm internal height by 90-mm internal diameter, hence is a good test of the applicability of our dielectric measurements. 

An important consideration is that once a cylindrical cavity is built, the cavity can be machined shorter in length to increase its target frequency, if the resonance is too low. Tuning the opposite direction, that is lowering a too-high frequency can be achieved by lengthening the cavity, but only if the end caps are designed to slide outwards smoothly.  In our design, this end-cap extension provides roughly 40~MHz of frequency reduction. It is clear that $\varepsilon'$ must be accurate to roughly $1 \%$ to ensure that such hybrid cavities can be accurately produced with this level of tunability.  

In light of this, we initially targeted a resonance of roughly 3.0~GHz with a height of 104 mm and 90 mm diameter, allowing for subsequent trimming modifications to match the target rubidium transition. From the data in Fig.~2 for $\varepsilon'$, we find that $\varepsilon'(3~\textrm{GHz}) = 2.03(4)$ and simulation of the COMSOL design resulted in a cavity frequency of 3.014 GHz.  After construction, our measured TE$_{011}$ mode was at 3.0141~GHz, allowing for careful shortening of the cavity to get in range of the tunability of the end-caps. Simulation of a shortened cavity height to 96.5~mm gave a resonance of 3.0439~GHz, where the final measurement of the constructed cavity yielded a frequency of 3.0449~GHz (within 0.03\% of our final target). The small tunability of the cavity end-caps then allows us to precisely match the 3.035732 GHz transition.

\section{Conclusion}
In this paper we report the dielectric constants and loaded-cavity loss tangents for three relevant machinable polymers in 3D microwave cavities, tracked over a large number of microwave resonances in the GHz range.  This allowed us to extract the frequency dependence of the dielectric constants, useful for designing of microwave cavities including these dielectric materials, as shown in a test case targeting a hyperfine resonance in rubidium.  Additionally, by comparison between finite element simulations that accounted for a variety of loss channels, we were able to extract loaded cavity loss tangents for these materials, which could influence material selections in future designs.  These measurements should aid researchers designing 3D microwave cavities for hybrid systems, especially those with limited tunability, or where the lowest loss must be attained.

\section{Acknowledgments}
This work was supported by the University of Alberta; the Natural Sciences and Engineering Research Council, Canada (Grants No. RGPIN-04523-16, No. RGPIN-2014-06618, and No. CREATE-495446-17);  the Alberta Quantum Major Innovation Fund; Alberta Innovates; and the Canada Research Chairs  (CRC) Program.

\section{Data Availability}
The data that supports the findings of this study are available within the article.

\end{document}